\documentclass[12pt]{article}

\input epsf
\usepackage{amssymb,amsmath,wrapfig}
\usepackage{cite}
\usepackage{graphicx}
\usepackage{slashed}
\usepackage{multirow}
\usepackage{float}
\usepackage{textcomp}
\usepackage{fancyhdr}
\usepackage{tikz}
\usepackage{feynmp}
\usepackage{axodraw4j}
\usepackage{pstricks}
\usepackage{color}

\restylefloat{table}

\makeatletter
\@addtoreset{equation}{section}
\makeatother

\begin{document}

        \begin{titlepage}

        \begin{center}

        \vskip .3in \noindent

        {\Huge \bf{The RG-improved Twin Higgs effective
         potential at NNLL}}

         \vspace{1.5cm}

        \large{Davide Greco\footnote{davide.greco@epfl.ch} and Kin Mimouni\footnote{kin.mimouni@epfl.ch}} \\

          \vspace{1cm}

       { \textit{Institut de Th\'{e}orie des Ph\'{e}nom\`{e}nes Physique,
        EPFL, Lausanne, Switzerland}\\
       \vspace{.1cm}
        }

        \vskip .5in
        {\bf Abstract }
        \vskip .1in
        
           \end{center}
We present the Renormalization Group improvement of the Twin Higgs effective potential at cubic order in logarithmic accuracy. We first introduce  a model-independent low-energy effective Lagrangian that captures both the pseudo-Nambu-Goldstone boson nature of the Higgs field and the twin light degrees of freedom charged under a copy of the Standard Model. We then apply the background field method to systematically re-sum all the one loop diagrams contributing to the potential. We show how this technique can be efficient to implicitly renormalize the higher-dimensional operators in the twin sector without classifying all of them. A prediction for the Higgs mass in the Twin Higgs model is derived and found to be of the order of $M_H \sim 120 ~\text{GeV}$ with an ultraviolet cut-off $m_*\sim 10-20 ~\text{TeV}$. Irrespective of any possible ultraviolet completion of the low-energy Lagrangian, the infrared degrees of freedom alone are  therefore enough to account for the observed value of the Higgs mass through running effects.

        \noindent

        \vfill
        \eject

        \tableofcontents

        \end{titlepage}

\section{Introduction} 
\label{sec:intro}

The principle of naturalness offers the main motivation for believing that new physics should exist slightly above the Electroweak Symmetry Breaking (EWSB) scale, $v \sim 246~ \text{GeV}$. In the Standard Model (SM), in fact, the Higgs boson is unstable under radiative corrections to its mass, so that it should be as heavy as the Planck scale, $M_P \sim 10^{19}$ $\text{GeV}$. This expectation is, however, of many orders of magnitude incompatible with the experimental results, that indicate the existence of a Higgs-like scalar resonance with mass $M_H \sim 125$ $\text{GeV}$. Natural considerations dictate therefore the existence of new physics at the TeV scale, which should be responsible for keeping the Higgs boson mass light. Concretely, weakly coupled supersymmetric scenarios or strongly interacting models enjoying the spontaneous breaking of a global symmetry are both candidates for a natural extension of the SM. In both cases, new light colored resonances should exist at the TeV scale, in the form of the supersymmetric stops \cite{Supersymmetry} or of the composite fermions coupled to the elementary quarks \cite{LightPartners}, which are required for $M_H$ to be predicted in the correct experimental range. The absence of any signal of these particles at the LHC is pushing these theories in the more fine-tuned regions and is forcing us to deeply reconsider the role of naturalness in the dynamics of the SM.

Another possible scenario which solves the hierarchy problem in a natural way and at the same time evades the constraints from direct searches at the LHC is however conceivable. It is given by the Twin Higgs paradigm \cite{Chacko:2005pe}: the new sector responsible for protecting the Higgs mass from large radiative corrections is given by a copy of the SM particles which is color-blind, namely it is not charged under the SM strong interactions. The new mirror partners which are required for the Higgs mass to be light are then invisible and cannot be detected at a hadronic collider. They are related to the SM fermions and bosons by a discrete $Z_2$ symmetry which, together with the spontaneous breaking of a global symmetry that turns the Higgs into a pseudo Nambu-Goldstone boson (pNGB), guarantees that the Higgs mass be insensitive to the UV contributions. The resulting possibility of having a natural EWSB with the absence of detectable new physics at the LHC has sparked interest in this class of models in the last years, \cite{ CompositeTwinHiggs, Tesi, Geller:2014kta, Chacko:2005vw, Chacko:2005un, SupTwin1, SupTwin2, Batra:2008jy, Craig:2014aea, Craig:2014roa, Burdman:2014zta, Craig:2015pha}, but many questions still remain open. In particular, an important problem is to analyze the capability of this scenario to reproduce the observed value of $M_H$, irrespective of any possible UV completion, supersymmetric or composite, of the low-energy Lagrangian. Since the Higgs mass is insensitive to the UV physics, in fact, the sole infrared (IR) degrees of freedom, namely the elementary SM particles and their mirrors, should be enough to account for the experimental indications. The Higgs mass receives then its most important contributions from the Renormalization Group (RG) evolution of the scalar potential from the UV down to the IR scale where $M_H$ is measured. Computing these running effects is crucial for an understanding of the feasibility of the Twin Higgs program as a new paradigm for physics at the electroweak (EW) scale.

In this paper, we study the Renormalization Group (RG) improvement of the Twin Higgs effective potential taking systematically into account the most important effects, due to QCD interactions and to loops of SM quarks and their twin copies. Our starting point will be a simplified low-energy effective Lagrangian that we write in full generality following the basic prescriptions of the Twin Higgs paradigm. These are the spontaneous breaking of a UV global symmetry and the existence of an extra elementary sector charged under a mirror of the SM gauge groups. The effective action is then simply given by the renormalizable SM interactions supplemented by two sets of higher-dimensional operators. The first set accounts for the non-linear Higgs interactions due to the pNGB nature of the Higgs scalar and it comprises the six-dimensional operators classified for instance in \cite{SILH2, Trott1, Trott2, Trott3}. The leading contribution to the potential generated by these latter is suppressed by the fine-tuning parameter $\xi=(v/f)^2$, where $f$ denotes the scale where the global symmetry is spontaneously broken. In presence of solely marginal and irrelevant interactions, in fact, the six-dimensional operators cannot renormalize the SM quartic coupling and mass parameter, but they can only  affect the running of other non-renormalizable operators with dimension $D \geq 6$. The effective potential must contain one operator of this type, $\mathcal{O}_6 = (H^\dagger H)^3$, where $H$ is the Higgs doublet. Its RG-evolution induced both by the linear and the non-linear interactions accounts for the contributions to the Higgs mass proportional to $\xi$. These effects are also common to any other natural extension of the SM with a pNGB Higgs in the spectrum. 

The second set of operators, specific to the Twin Higgs construction, describes the interactions between the Higgs boson and the twin fermions. Its most distinctive feature is the existence of a relevant term with dimension $D=3$, namely the twin quark mass parameter, that is generated before EWSB, \cite{CompositeTwinHiggs}. Together with this latter, a series of non-renormalizable operators must be taken into account, whose leading contribution to the potential is not necessarily proportional to $\xi$, unlike the case of the six-dimensional operators made up of SM fields only. Due to the super-renormalizable mass term, in fact, the higher-dimensional interactions in the Twin sector can not only affect the running of other irrelevant operators with $D>4$, but they can also renormalize the SM quartic coupling and mass term. If we consider, for instance, the dimension-five operator $\mathcal{O}_5=(H^\dagger H)\bar{\widetilde{q}}~ \widetilde{q}$, with $\widetilde{q}$ a twin quark, we can easily construct a one-loop diagram contributing to the running of the quartic coupling. If two vertices are given by $\mathcal{O}_5$, two insertions of the twin quark mass are enough to generate a marginal operator. Similar considerations are valid for the other higher-dimensional operators, which can always renormalize the lower-dimensional ones through the insertion of an increasing number of the relevant three-dimensional interaction. In particular, we would need to classify all the non-renormalizable operators in the twin sector up to dimension $D=9$ in order to fully capture the correction to the Higgs mass up to the order $\xi$. As a consequence, a diagrammatic computation of the RG-evolution of the effective potential results to be quite complicated, since no existing classification of the Twin non-renormalizable operators exists. Moreover, the number of diagrams one has to compute to renormalize the quartic coupling and $\mathcal{O}_6$ is big enough to discourage the usage of this diagrammatic approach.

It is possible to avoid the full classification of the operators in the Twin sector by making use of a more clever technique to compute the Higgs effective potential, the background field method. As it is well known, this procedure allows to derive the RG-improved action automatically re-summing a whole series of diagrams and without needing to calculate all the single operators that are renormalized along the RG flow. If this method may be just an alternative in the SM, for the Twin Higgs model it provides instead the fastest way to calculate the contribution of the extra light degrees of freedom. We will therefore derive our expression for the Higgs mass using the background field method. The result will be organized as an expansion in logarithms, as usual, and we will show how to systematically include all the contributions to the effective potential that are generated along the flow as higher powers in the logarithmic series are included. We will renormalize the effective action up to the third order in the expansion parameter, classifying and discussing separately the leading contribution, the quadratic correction and finally the cubic expression for the Higgs mass. 

This paper is organized as follows. In Section \ref{sec:Review}, we will review the Twin Higgs paradigm and write down its effective low-energy Lagrangian. After briefly recalling the leading result for the effective potential, in Section \ref{sec:NLL} we will apply the background field method to the Twin Higgs model and show how to derive the RG-improved effective potential at quadratic order. In Section \ref{sec:NNLL}, we shall extend the computation to include the cubic terms. Section \ref{sec:Results} contains a discussion of the final results, the validity of our approximation and the prediction for the Higgs mass that we get in the Twin Higgs model. In particular, Figs.~(\ref{fig:HiggsMass}) and (\ref{fig:HiggsMass2}) represent the most important result of this work and contain the numerical estimation of $M_H$ both in the SM and in its Twin extension. We conclude summarizing our findings in Section \ref{sec:Conclusions}.


\section{The Twin Higgs low-energy Lagrangian} 
\label{sec:Review}

The Twin Higgs paradigm is an interesting alternative to theories which conceive the Higgs scalar as a pNGB, like for instance the Composite or the Little Higgs \cite{MinimalCompositeHiggs,LittleHiggs}. Two are the basic assumptions of any realization of this scenario \cite{Chacko:2005pe}. First of all, at a generic UV scale $m_*$ there must exist some extension of the SM whose Higgs sector enjoys an approximate global symmetry, $G$. This latter is spontaneously broken at an IR scale $f$ to some unbroken subgroup $H$ so that seven Goldstone bosons (GB) are delivered in the spectrum; four of them are identified as the Higgs doublet. The second element is an approximate discrete $Z_2$ symmetry that interchanges in the UV every SM particle with a corresponding mirror particle charged under a twin copy of the SM gauge groups, $\widetilde{\text{SM}}$. 

The mechanism that allows a natural EWSB employs the explicit breaking of both these symmetries. The weak and electromagnetic interactions together with the Yukawa couplings violate, in fact, the global symmetry $G$. As a result, three of the seven GB's are eaten to give mass to the twin gauge bosons, a potential for the Higgs doublet is generated and the Higgs scalar is turned into a pNGB. An exact discrete symmetry, on the other hand, guarantees that the mass term in the Higgs potential be trivially invariant under $G$, so that it does not contribute a physical mass to the GB's. These latter are then completely insensitive to any quadratic contribution proportional to $m_*^2$ and originated by loops of heavy particles or by the high-energy propagation of the light degrees of freedom. The $G$-breaking terms in the potential, like the Higgs quartic coupling, are at most only logarithmically sensitive to the scale $m_*$ and must be proportional to $g^4$ and $y^4$, where $g$ collectively indicate the weak gauge couplings and $y$ the Yukawas. An explicit soft breaking of the $Z_2$ symmetry is however necessary to generate a small quadratic mass term that in turn allows a tunable minimum of the potential to exist. Therefore the discrete symmetry, while being potentially respected by all the SM and $\widetilde{\text{SM}}$ interactions, must be softly broken by some UV effects. A natural hierarchy between the EW scale $v$ and the GB decay constant $f$ is generated without requiring the existence of new light particles charged under the SM. The UV scale $m_*$, where the heavy fields with SM quantum numbers reside, can thus be pushed up to $m_* \sim 10 ~\text{TeV}$, out of the LHC reach, without in any way worsening the tuning between $v$ and $f$.

The Higgs effective potential being insensitive to the UV scale, it is crucial to study how it is affected by the IR physics. In particular, it is important to derive an expression for the Higgs boson mass and understand how light it can be, also in comparison with its experimental value. To tackle these questions, we aim at analyzing the RG-improvement of the effective potential including the running of the quartic coupling induced by the light degrees of freedom present in the Twin Higgs paradigm. Our starting point is the low-energy Lagrangian at the scale $m_*$ generated after integrating out the UV physics together with the heavy mirror copy of the Higgs doublet. We consider a non-linear implementation of the Twin Higgs symmetries, so that also the radial mode of the linear realization is integrated out. We will be completely agnostic as regards the particular UV completion of the theory, which could be a strongly interacting composite dynamics \cite{CompositeTwinHiggs, Tesi}, a weakly coupled supersymmetric sector \cite{SupTwin1,SupTwin2} or the linear model itself, and as regards any possible UV mechanism that softly breaks the discrete symmetry. At the same time, we will not specify any particular symmetry breaking coset; as long as it delivers seven GB's, it could be $SU(4)/SU(3)$ as in the original model \cite{Chacko:2005pe} or $SO(8)/SO(7)$ as in the minimal composite UV completion \cite{CompositeTwinHiggs, Tesi}. We will also neglect the tree-level contribution of all the higher-dimensional operators, like current-current or four fermions operators, that could be originated after integrating out heavy bosonic or fermionic resonances. Their Wilson coefficients at the scale $m_*$ are in fact model-dependent and moreover they are suppressed both by the weak coupling between the light degrees of freedom and the new dynamics and by inverse powers of $m_*$. Supposing this scale to be in the multi-TeV range, as in the spirit of the Twin Higgs paradigm, the initial conditions for these type of higher-dimensional operators can be safely taken to be zero. Our Lagrangian will however take into account the two basic elements of the twin Higgs construction, namely the presence of non-linear Higgs interactions due to the pNGB nature of the Higgs boson and the existence of extra light degrees of freedom charged under the $\widetilde{\text{SM}}$. The remaining non-renormalizable terms that we neglected at the tree-level will be seeded at one-loop by the non-linear Higgs dynamics. We will consider just the most relevant contributions to the potential, originating from the $G$-breaking gauge and Yukawa interactions. We neglect the weak gauge couplings, whose effects are much smaller than those in the quark sector, and  we keep only the terms proportional to the top Yukawa coupling, which generates the most important corrections to the potential. Under all these assumptions, the effective Lagrangian at the scale $m_*$ is:
\begin{equation}\label{LagInitial}
\begin{array}{ll}
\mathcal{L}(m_*) =& \displaystyle  (D^\mu H^\dagger)( D_\mu H ) - V(H^\dagger H, m_*)~+\\ 
& \displaystyle  \bar{Q}_L i \slashed D Q_L  + \bar{t}_R i \slashed D t_R - {y_t(m_*)}  \left[ f ~\bar{Q}_L {H'\over \sqrt{2 H^\dagger H}} \sin \left( {\sqrt{2H^\dagger H} \over f}\right) t_R ~+ \text{h.c.} \right]+\\
& \displaystyle \overline{\widetilde{t}} ~i \slashed D ~\widetilde{t} - {\widetilde{y}_t(m_*) \over \sqrt{2}}f \cos\left({\sqrt{2 H^\dagger H}\over f} \right)\overline{\widetilde{t}}~ \widetilde{t} .
\end{array}
\end{equation}
In the previous equation, $y_t$ and $\widetilde{y}_t$ denote the SM top Yukawa coupling and its twin; they are initially equal due to the approximate $Z_2$ symmetry: $y_t(m_*) = \widetilde{y}_t (m_*)$. The twin tops $\widetilde{t}$ are not charged under the SM and therefore do not form any doublet with the corresponding twin bottom. This latter can then be neglected since its contribution to the RG flow of the Higgs potential would be proportional to $\widetilde{y}_b$ and is thus sub-leading. The covariant derivatives of the fermion fields contain the strong interactions with coupling $g_S$ for the $SU(3)$ SM gauge groups and $\widetilde{g}_S$ for its twin. Because of the twin symmetry, we have again $g_S(m_*) = \widetilde{g}_S(m_*)$. $H$ is instead the SM Higgs doublet,
\begin{equation}\label{Doublet}
H = {1\over \sqrt{2}} \left(
\begin{array}{ll}
\pi _1 + i \pi_2\\
h + i \pi_3
\end{array}\right);
\end{equation}
we define $H'= i \sigma_2 H^*$ and $V(H^\dagger H, m_*)$ is the Higgs effective potential at the scale $m_*$:
\begin{equation}\label{HiggsPotential}
V(H^\dagger H, m_*) = L(m_*) \sin^2\left({\sqrt{2H^\dagger H}\over f}\right) + F(m_*) \left[ \sin^4\left({\sqrt{2H^\dagger H}\over f}\right)+ \cos^4\left({\sqrt{2H^\dagger H}\over f}\right) \right].
\end{equation}
The mass term $L$ is generated by the $Z_2$ breaking interactions, whereas the function $F$ arises at the tree-level after integrating out the UV sector; their explicit form at $m_*$ is model-dependent and provides an $O(1)$ initial condition for the running of the effective potential. 

The low-energy Lagrangian fully takes into account the pNGB nature of the Higgs scalar by introducing the non-linear trigonometric interactions between the Higgs doublet and fermions. The effective potential has also the specific trigonometric dependence that is dictated by the existence of a non-linearly realized spontaneous symmetry breaking coset. It is convenient to make a field redefinition in order to recover the SM Lagrangian supplemented by higher-dimensional operators and to simplify the initial conditions at the scale $m_*$ for the relevant Wilson coefficients. We therefore redefine the Higgs doublet as
\begin{equation}\label{FieldRedef}
H \rightarrow f {H\over \sqrt{2H^\dagger H}} \sin \left({\sqrt{2H^\dagger H}\over f}\right)
\end{equation}
and recast the Lagrangian in Eq.~(\ref{LagInitial}) in the following form:
\begin{equation}\label{LagMstar}
\begin{array}{ll}
\mathcal{L}(m_*) = & \displaystyle (D^\mu H^\dagger)( D_\mu H) + {1\over 2 f^2}\left [c_H(m_*) + d_H(m_*) { H^\dagger H\over 4 f^2} \right] \mathcal{O}_H+{c_H'(m_*)\over f^2}\mathcal{O}_H'- V(H^\dagger H,m_*) +\\
& \displaystyle \bar{Q}_L i \slashed D Q_L + \bar{t}_R i \slashed D t_R + \overline{\widetilde{t}}~ i\slashed D ~\widetilde{t} -\left[ {y_t(m_*)} \bar{Q}_L H' t_R + {\widetilde{y}_t (m_*) f\over \sqrt{2}} \sqrt{1- {2H^\dagger H\over f^2}}~ \overline{\widetilde{t}} ~ \widetilde{t}+\text{h.c.}\right],
\end{array}
\end{equation}
where the potential can now be written as
\begin{equation}\label{VPot}
V(H^\dagger H,m_*) = 2 \mu^2(m_*) H^\dagger H+ 4 \lambda(m_*) (H^\dagger H)^2+8 {c_6(m_*)\over f^2} \mathcal{O}_6.
\end{equation}
Using the notation of \cite{SILH2}, we have introduced the dimension-6 operators $\mathcal{O}_H= \partial^\mu (H^\dagger H) \partial_\mu (H^\dagger H) $ and $\mathcal{O}_H'= H^\dagger H (D^\mu H^\dagger)( D_\mu H)$. It is straightforward to verify that $c_H(m_*) = 1$, whereas $O_H'$ is not generated at the tree-level with our choice of basis, $c_H'(m_*) = 0$.\footnote{Notice that $\mathcal{O}_H'$ corrects the $W$ boson mass at order $\xi$, whereas in the basis (\ref{FieldRedef}) no correction to the gauge boson masses is induced. We did not report the low-energy Lagrangian in the gauge sector, but it can be found in \cite{CompositeTwinHiggs}, for instance. As a consequence, this operator is absent at the tree-level. For the same reason, the eight-dimensional companion operator of $\mathcal{O}_D$, $\mathcal{O}_D'= (H^\dagger H)^2 |D_\mu H^\dagger|^2$, has vanishing boundary conditions when matching with the Twin Higgs Lagrangian in our basis. Since only the tree-level initial conditions for the eight-dimensional operators can affect the RG-improvement of the potential at cubic order, we can completely neglect $\mathcal{O}_D'$ from our Lagrangian.} Only the RG-evolution will seed this operator at loop-level. Notice also the presence of the dimension-8 operator $\mathcal{O}_D = H^\dagger H \partial^\mu (H^\dagger H) \partial_\mu (H^\dagger H)$ , with $d_H(m_*)=8$, which is necessary to capture all the effects due to the running in the Twin sector, as we shall see. The Wilson coefficients in the Higgs potential can be expressed as functions of $L$ and $F$ at the scale $m_*$, although the explicit relation is not relevant for the analysis of the IR contributions to the Higgs mass. However, one can check that the initial condition for $c_6$ is simply $c_6(m_*) = 0$, so that the operator $\mathcal{O}_6 =  (H^\dagger H)^3$ is generated only through the running.  All the contributions to the Higgs mass or to other observables due to the higher-dimensional operators in the SM sector are suppressed by powers of $\xi$, which measures the degree of tuning between the EW scale and the GB decay constant. The parameter $\xi$ is also constrained to be small by electroweak precision tests (EWPT) which set a bound $\xi \leq 0.2$. As regards the Twin sector, notice finally that the non-renormalizable interactions generated at the tree-level are all collected in the function of the Higgs field which accompanies $\widetilde{y}_t$. From the Lagrangian in Eq.~(\ref{LagMstar}), we can also derive the expressions of the top masses and their scale separation. After EWSB, we have in fact $m_t = y_t v/ \sqrt{2}$ for the SM tops and $\widetilde{m}_t = \widetilde{y}_t f \sqrt{1-\xi}/\sqrt{2}$ for their twins.  

\subsection{The Higgs mass and the LL result}

The potential at the scale $m_*$ gives rise to a first small UV contribution to the Higgs mass. This is a model-dependent tree-level effect that arises after integrating out the heavy physics. We have:
\begin{equation}\label{MHUV}
(M_H^2) _ {UV} \sim  \lambda(m_*) v^2.
\end{equation}
The RG evolution of the potential induced by the light degrees of freedom generates other log-enhanced IR corrections due to the running from $m_*$ down to the low-energy scale where the Higgs mass is experimentally measured, for instance $m_t$, the top mass scale. The Higgs mass receives then a second contribution, $(M_H^2)_{IR}$, which is model-independent and proper to any possible UV completion of the Twin Higgs paradigm. Our full prediction for this observable is therefore:
\begin{equation}\label{HiggsMass}
M_H^2 = (M_H^2)_{UV}+(M_H^2)_{IR},
\end{equation}
where $(M_H^2)_{IR}$ can be expressed at a generic renormalization scale $\mu$ as a function of the renormalized Wilson coefficients appearing in Eq.~(\ref{LagMstar}). At first order in $\xi$, we have:
\begin{equation}\label{HiggsMassFormula}
(M_H^2)_{IR} (\mu) = 8 \left[\lambda(\mu) + 3 ~ c_6(\mu) ~\xi   \right]\left[1- \left(c_H(\mu) +c_H'(\mu)\right) ~\xi \right]~v^2. 
\end{equation}
Once the RG flow to the IR scale has been computed to the desired level of accuracy, one can match with the UV mass term so as to reproduce the observed value of the Higgs mass, $(M_H^2)_{Exp} = (125 ~\text{GeV})^2$. We aim at deriving an expression for the IR RG evolution in order to judge how important the running effects are and to analyze which value of the UV threshold correction is more suitable. This will in turn give information on what kind of UV completion can be imagined to generate $(M_H^2)_{UV}$ of the right size.

The computation of the RG evolution of the Higgs potential can be carried out at different orders in an expansion in logarithms. The leading contribution is obtained by neglecting the running of the top Yukawas and the strong couplings and retaining only the first power in the logarithmic expansion. We call this order leading logarithm (LL) result. Using the standard Coleman-Weinberg technique, one finds that only $\lambda$ can be generated at the leading order, whereas $c_6$ is still vanishing; the Higgs mass is then \cite{CompositeTwinHiggs}:
\begin{equation}\label{MHIRLL}
(M_H^2)_{IR}^{LL}(m_t) = {3\over 8\pi^2}\left[
 y_t^4(m_*)  \log\left( m_*^2 \over m_t^2\right) + \widetilde{y}^4_t(m_*) \log\left( m_*^2 \over \widetilde{m}_t^2\right) \right]  \left(1- \xi\right) v^2,
\end{equation}
which is the sum of two different contributions. The first one is proportional to $y_t^4$ and is induced by the running of the quartic coupling due to loops of SM fermions, while the second is of order $\widetilde{y}_t^4$ and results from analogous loops of twin tops. Notice also that we have included the first correction to the leading logarithm, proportional to $\xi$. This effect is usually smaller and parametrically belongs to the next class of contributions. By setting $y_t$ to the experimental value at the scale $m_t$, we can estimate the value of the Higgs mass generated by the IR physics. For $\xi =0.1$ and $m_* = 10 ~\text{TeV}$, we predict $(M_H)_{IR} \sim 150 ~\text{GeV}$, which is far above the experimental observations. A more accurate analysis that takes into account the running of the Yukawas, the strong couplings and the higher-dimensional operators can drastically change this prediction and the consequent necessary size of the UV threshold correction. 

In this paper we will study the RG-improvement of the potential and derive the first two corrections of the LL Higgs mass, up to effects that are cubic in the logarithmic series. Indicating with $t= \log(m_*^2/ \mu^2)$ the expansion parameter, where $\mu$ is again the renormalization scale, we shall consider first of all the next-to-leading logarithmic contribution to the potential (NLL), which incorporates all the effects proportional to $t^2$. We will include in this class also the smaller $\xi ~t^2$ contributions to the Higgs mass, that would belong to the next class of corrections; for simplicity of exposition we classify them in the same category as the other $t^2$ terms. We will neglect all the other powers of $\xi$, which are much smaller due to the constraint from EWPT. The second correction we shall compute is the next-to-NLL (NNLL), which contains only the $t^3$ effects. We will not compute the smaller $\xi ~ t^3$ corrections, which are part of the next class of contributions.

\section{The NLL effective potential}
\label{sec:NLL}

The RG-improvement of the Higgs effective potential is the result of all the physical effects that induce an evolution of the Wilson coefficients when changing the energy scale of a process. While running down from $m_*$ to $m_{t}$, the high energy - or equivalently short distance - degrees of freedom are integrated out and the initial parameters in the Lagrangian must be redefined to properly describe the physics at low-energy and to eliminate the loop divergences. In particular, in order to fully capture the NLL corrections to the potential, we have to take into account three important effects. First of all, the top Yukawa couplings in the SM and $\widetilde{\text{SM}}$ sectors evolve along the RG flow because of the strong interactions and the coupling with the Higgs field. The adequate inclusion of this running contributes to the potential at order $t^2$. Secondly, the dimension-6 operators $\mathcal{O}_H$ and $\mathcal{O}_H'$ are corrected with respect to their tree-level initial values due to loops of fermions, thus affecting  the Higgs mass at order $\xi t^2$. Finally, also the Higgs wave function receives a non-vanishing correction from top loops resulting in a non-canonical scalar field; the wave function renormalization will affect the whole NLL result, both at $t^2$ and $\xi t^2$ level.

We will derive the NLL effective potential using the background field method, as developed in standard textbooks of quantum field theory \cite{Weinberg, Peskin}. This technique proves  to be extremely powerful for theories like the Twin Higgs model while at the same time being perfectly equivalent to the diagrammatic approach. Due to the presence of non-renormalizable interactions, in fact, new operators are generated along the RG-flow at each step of the running, so that using a more conventional diagrammatic procedure one would need to keep track of all them and compute an increasing number of diagrams. The application of the background technique, instead, treats the Higgs field as an external spectator and re-sums automatically a huge class of diagrams without much increasing the effort as more powers of $t$ are included. At the quadratic level, this method is so powerful that the sole renormalization of the twin top propagator is equivalent to the computation of an order of ten loops with the diagrammatic approach. We shall devote this Section to the presentation of the background field method and its usage to derive a general RG-improved Coleman-Weinberg formula for the effective potential. This latter will be applied to the Twin Higgs Lagrangian in order to compute the NLL correction to the Higgs mass. 

\subsection{The background field}

The background field method is based on the idea that one can explicitly integrate out the short distance degrees of freedom after separating them from the low-energy modes. Since we are interested in computing the effective potential for the Higgs boson, our starting point is to split the scalar doublet in two parts, a background spectator field and a quantum fluctuation:
\begin{equation}\label{Separ}
H = H_c + \widehat{\eta}.
\end{equation}
$H_c$ indicates the classical field configuration for the Higgs doublet; it comprises all the low-energy modes that we will keep in the spectrum and for which we will find a potential. $\widehat{\eta}$ denotes instead the dynamical fluctuations over the classical field; these are the high-energy modes we seek to integrate out. After separating the short distance from the large distance degrees of freedom, we can recast the top and twin top sectors of the Lagrangian in Eq~(\ref{LagMstar}) as follows:
\begin{equation}\label{LagF}
\mathcal{L}^F(m_*) =\mathcal{L}^F_{Kin}(m_*)-\bar{Q}_L m(H_c)' t_R  - {\widehat{y}_t(H_c) }\bar{Q}_L \widehat{\eta}^{~'} t_R-\widetilde{m}_t(H_c)~ \bar{\widetilde{t}}~\widetilde{t}+{\widehat{\widetilde{y}}_t(H_c)^\dagger~ \widehat{\eta}\over \sqrt{2}}~\bar{\widetilde{t}} ~\widetilde{t}+\text{h.c.},
\end{equation}
where $\mathcal{L}^F_{Kin}(m_*)$ collectively indicates the kinetic terms of the fermion fields. Expanding the Lagrangian in powers of $\widehat{\eta}$, we kept only the linear interactions of the high-frequency modes with the top quarks, since the remaining non-linear interactions do not contribute at the NLL order. The coupling between the $\widehat{\eta}$ fields and fermions is in general a background-dependent function; in the SM, it is trivially equivalent to the top Yukawa, but in the Twin sector it is has a specific functional form. Promoting the Yukawa couplings to spurions of the spectator Higgs field, we have introduced the following background-dependent quantities:
\begin{equation}\label{Yukawas}
\widehat{y}_t (H_c) \equiv y_t , \qquad \widehat{\widetilde{y}}_t (H_c) \equiv \widetilde{y}_t ~{H_c \over f}{1 \over \sqrt{1- {2H^\dagger_c H_c\over f^2}}}.
\end{equation}
Also the fermion masses at the tree-level can be considered as functions of the spectator $H_c$ and treated formally as spurions; one easily finds:
\begin{equation}\label{MassesBackground}
m_t (H_c) = y_t  H_c, \qquad  \widetilde{m}_t(H_c) = {\widetilde{y}_t f \over \sqrt{2}} \sqrt{1-{2H_c^\dagger H_c \over f^2}}.
\end{equation}
The physical value of the mass parameters is obtained by setting the background doublet to its EW vacuum expectation value, thus recovering the standard expressions.\footnote{The top Yukawas in Eqs.~(\ref{MassesBackground}) and (\ref{Yukawas}) are both evaluated at the scale $m_*$; from now on, we will omit to specify the scale where the initial condition of the bare parameters originates, unless differently stated they will all be considered at the cut-off.} 

Let us now consider the scalar sector of the theory. After separating the short distance modes from the long-distance ones, a set of new interactions between the background field and the quantum fluctuation is generated. Of these, only a few are relevant for the NLL computation; in practice, we just have to take into account that the kinetic term for $\widehat{\eta}$ becomes non-canonical and acquires a background dependence. We have in fact:
\begin{equation}\label{LagS}
\mathcal{L}^S(m_*) \supset |D^\mu H_c|^2 +{1\over 2 f^2} \left(c_H + d_H {H_c^\dagger H_c \over 4 f^2} \right) \mathcal{O}_H(H_c) + Z_{\widehat{\eta}} (H_c)|D^\mu ~\widehat{\eta}|^2,
\end{equation}
with 
\begin{equation}\label{ZEta}
Z_{\widehat{\eta}}(H_c) = 1+ 2 c_H {H_c^\dagger H_c \over f^2}+  d_H {(H_c^\dagger H_c)^2\over 2 f^4}.
\end{equation}
As for the fermionic sector, the previous equation serves as an initial condition for the wave function of the high-energy modes, which will be modified along the flow by quantum corrections. One could choose to perform a proper field redefinition in order to eliminate the background dependence and render the fluctuation canonical. We will work, instead, with a non-canonical basis and integrate out the high-energy degrees freedom without redefining the $\widehat{\eta}$ fields. As a consequence, we will have to write down a separate evolution equation for the wave function which will be coupled to the $\beta$-functions of the Yukawa couplings. Despite this additional feature, choosing a non-canonical basis has many advantages and allows to efficiently re-sum all the diagrams generated by insertions of the higher-dimensional operators $\mathcal{O}_H$, $\mathcal{O}_H'$ and $\mathcal{O}_D$. Only after deriving the effective potential will we perform the field redefinition and find the Higgs mass in the canonical basis.
 
\subsection{$\beta$-functions in the Higgs background}

After separating the quantum fluctuation from the long distance modes and finding the background-dependent couplings and fermion masses, the short distance degrees of freedom must be integrated out to derive the effective action at low energies. In the scalar sector, this process generates a quantum contribution to the wave function of $H_c$ and $\widehat{\eta}$ and also renormalizes $c_H$ and $c_H'$. In the fermionic sector, the integration of the high-frequency modes results in the redefinition of the background-dependent Yukawa couplings and masses, which start evolving with the energy scale. In this section, we will derive a set of coupled differential equations for the Higgs wave function and the Yukawas in the classical background. They are the generalization of the usual $\beta$-functions for a general theory with a non-canonical Higgs and field-dependent Wilson coefficients.

\begin{figure}[t!]
\begin{center}
\includegraphics[width=0.9\textwidth] {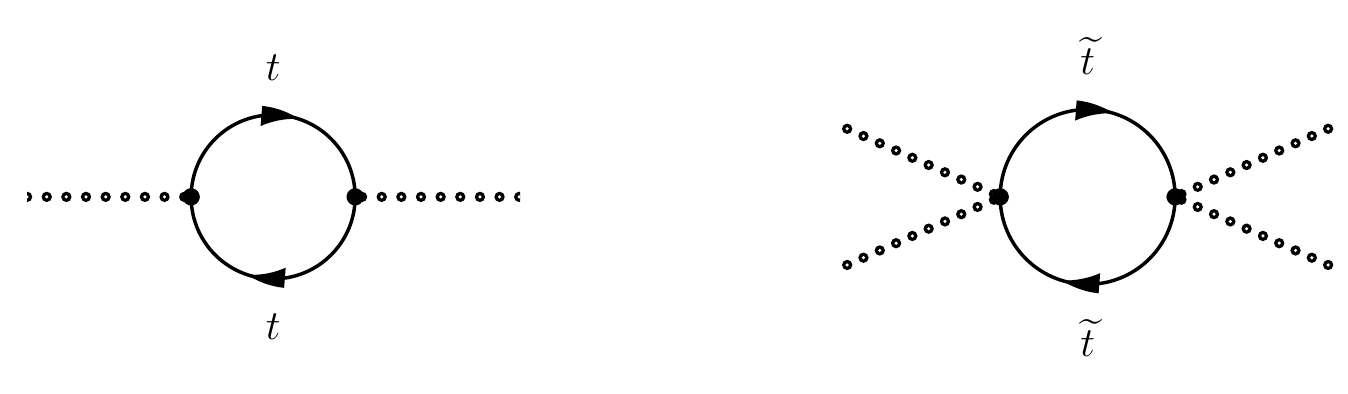}
\end{center}
\caption{One loop diagrams contributing to the wave function renormalization (on the left) and to the running of $c_H'$ (on the right). The external dotted lines denote the background field $H_c$. }
\label{fig:WaveFunction}
\end{figure}

We start our study with the scalar sector. The running of the wave function and of the other Lagrangian parameters is induced in this case by loops of fermions; one would formally need to split also the top fields into long distance and short distance modes and integrate out these latter. This is completely equivalent to computing the one-loop diagrams in Fig.~(\ref{fig:WaveFunction}) with $N_c=3$ colors circulating for both SM and $\widetilde{\text{SM}}$ quarks. The coupling between the background field and the fermionic fluctuation is obtained by expanding the mass terms in Eq.~(\ref{LagF}) in powers of the spectator $H_c$. For the SM, only the usual linear coupling proportional to $y_t$ exists and therefore loops of tops can only renormalize the wave function of the Higgs. For the Twin sector, instead, the first non-trivial coupling is quadratic in the Higgs background, so that no contribution to the wave function can be obtained from the mirror tops. One-loop diagrams of twin fermions will however renormalize the higher-dimensional operator $\mathcal{O}_H'$. At first order in the expansion parameter $t$, the Lagrangian at a generic renormalization scale $\mu$ becomes:
\begin{equation}\label{LagSRen}
\mathcal{L}(t)^S \supset Z_H(t)|D^\mu H_c|^2 + Z_{\widehat{\eta}} (H_c,t) |D^\mu ~ \widehat{\eta}|^2 +{c_H(t)\over 2 f^2} \mathcal{O}_H + {c_H'(t)\over f^2} \mathcal{O}_H'+ {d_H \over 8 f^4} \mathcal{O}_D,
\end{equation}
where
\begin{equation}\label{ZRen}
\begin{array}{ll}
\displaystyle Z_H(t) = 1+{N_c y_t ^2 \over 16 \pi^2} t ,\quad c_H (t)= c_H, \quad c_H'(t)= {N_c \widetilde{y}_t^2 \over 16 \pi^2} t, \vspace{0.2cm}\\
\displaystyle Z_{\widehat{\eta}} (H_c, t) =  Z_H(t) + (2 c_H(t) + c_H'(t)){H_c^\dagger H_c \over f^2}+  d_H {(H_c^\dagger H_c)^2\over 2 f^4}.
\end{array}
\end{equation}
The one-loop integration of the high-energy fermionic modes also induces a renormalization of $d_H$ which however we can neglect for our purposes. Only the tree-level value of this parameter contributes to the Higgs effective potential at the NLL because $\mathcal{O}_D$ can only renormalize $\mathcal{O}_6$ which in turn can be first generated at order $t^2$. 

\begin{figure}[t!]
\begin{center}
\includegraphics[width=0.9\textwidth] {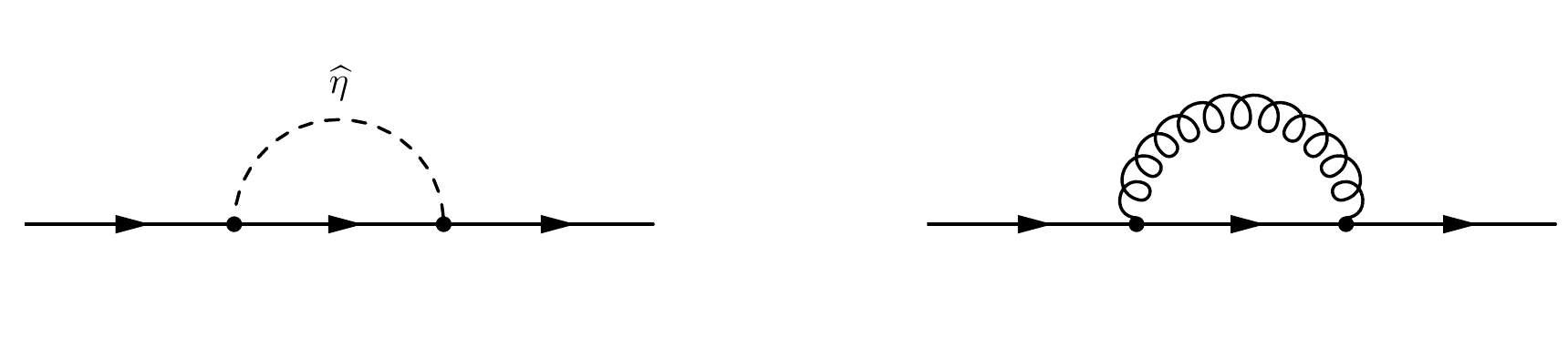}
\end{center}
\caption{One-loop diagrams displaying the renormalization of the top quark propagator due to the interaction with the Higgs quantum fluctuations (on the left) and with gluons (on the right). The solid black lines denote the fermion field, either the SM tops or their $\widetilde{\text{SM}}$ mirrors, whereas the curly line stands both for the $SU(3)$ and the $\widetilde{SU}(3)$ gluons. The dashed line stands for the quantum fluctuation.}
\label{fig:FermProp}
\end{figure}

Let us now consider the fermionic sector of the Twin Higgs theory. The process of integrating out the high energy modes of the Higgs field translates in this case into a renormalization of the top quarks propagator, as in Fig.~(\ref{fig:FermProp}). Together with the scalar fluctuations, a contribution to the running of the Yukawas is also generated by QCD gluons, both in the SM and in the $\widetilde{\text{SM}}$. The computation of these effects is standard and leads to a background-dependent quantum correction to the quarks wave functions and their mass. After rescaling the fermion fields, we find:
\begin{equation}\label{YRen}
y_t (t) = y_t + {y_t\over 64 \pi^2} \left( 16 g_S^{~2} - 3{ \widehat{y}^{~2}_t(H_c)\over Z_{\widehat{\eta}}(H_c) } \right) t ,  \qquad \widetilde{y}_t (t) = \widetilde{y}_t + {\widetilde{y}_t\over 64 \pi^2} \left( 16 \widetilde{g}_S^2 - 3{ \widehat{\widetilde{y}}^{~2}_t(H_c)\over Z_{\widehat{\eta}}(H_c) } \right) t.
\end{equation}
Since the scalar fluctuation is still non-canonical in our basis, every propagator of the $\widehat{\eta}$ fields is accompanied by an inverse power of $Z_{\widehat{\eta}}(H_c)$, which in turn must appear explicitly in the evolution of the Yukawa couplings. This is why it is convenient to keep the short distance modes non-canonical: all the contributions to the running proportional to $c_H$, $c_H'$ and $d_H$ will be automatically re-summed in the denominator of the beta functions without any need of computing additional diagrams. The sole renormalization of the top quark propagator in the background field language is enough to consistently keep track of all the higher-dimensional operators that will be generated along the flow. 

The RG evolution of the Yukawa couplings and of the Higgs wave functions can be elegantly described by a set of background-dependent coupled differential equations that take into account the physical effects we have encountered so far. These $\beta$-functions will re-sum all the leading logarithms in the energy flow; for a general Wilson coefficient $c$ they can be defined as:
\begin{equation}
\beta_c = {d c(t) \over dt}.
\end{equation}
From the previous results, we then easily find the following RG-equations:
\begin{equation}\label{BGBetaFunctions}
\begin{array}{ll}
\displaystyle \beta_{{y} _t} = { y_t(H_c, t) \over 64 \pi^2} \left( 16 g_S^2(t) - 3 {y_t^2(H_c,t)\over Z_{\widehat{\eta}}(H_c,t)}\right), \quad \beta_{Z_{\widehat{\eta}}} = {3 y_t^2(H_c,t)\over 16\pi^2}+ {3\widetilde{y}_t^2(H_c,t)\over 16 \pi^2} { H_c^\dagger H_c\over f^2}, \vspace{0.2cm} \\ 
\displaystyle \beta_{\widetilde{y} _t}= { \widetilde{y}_t(H_c,t) \over 64 \pi^2} \left( 16 \widetilde{g}_S^2(t) - 3 {\widetilde{y}^2_t(H_c,t)  \over Z_{\widehat{\eta}}(H_c,t)} {2 H_c^\dagger H_c\over f^2}{1\over {1-{2H^\dagger_c H_c\over f^2}}}\right), \quad  \beta_{Z_H} = {3 y_t^2(H_c,t)\over 16\pi^2}\bigg |_{H_c=0}.
\end{array}
\end{equation}
The $\beta$-functions we have just derived are valid in a non-canonical basis; once they are solved, we need to redefine the background field $H_c$ in order to compute the RG-improved physical quantities. For instance, the SM top Yukawa in the canonical basis is obtained with the simple combination
\begin{equation}\label{YPhys}
y_t^{Phys} (t )= {y_t(0,t) \over \sqrt{Z_H(t)}},
\end{equation}
with $y_t(0,t)$ being the running Yukawa coupling evaluated at zero spectator field; analogous relations hold for the remaining parameters. Notice that we do not need an explicit $\beta$-function for $c_H$ and $c_H'$ since their RG-evolution is already absorbed in the running of the wave function for the fluctuation $\widehat{\eta}$. This is another reason why it is advantageous to keep the Higgs field non-canonical. Finally, the running of the top quark masses, which are the quantities we will need in the Coleman-Weinberg formula, is directly related to the evolution of the Yukawas. We have:
\begin{equation}\label{MassesRen}
m_t(H_c, t) = y_t(H_c, t) H_c, \qquad \widetilde{m}_t (H_c,t) ={\widetilde{y}_t(H_c,t) f \over \sqrt{2}}\sqrt{1-{2 H_c^\dagger H_c\over f^2}},
\end{equation}
where $y_t(H_c,t)$ and $\widetilde{y}_t(H_c,t)$ denote the solution of the $\beta$-functions in the Higgs background. This is the starting point for the computation of the RG-improved effective potential.

\subsection{RG-improved Coleman-Weinberg formula and Higgs mass}
 
The Coleman-Weinberg procedure to compute the effective potential is an efficient way of re-summing all the one-loop diagrams contributing to the low-energy action with a generic number of external scalar legs. This formally corresponds to calculate the vacuum energy, or cosmological constant, of the theory in an external background. In order to improve the LL result and include all the leading logarithms that are generated during the running, we can use an evolution equation for the cosmological constant itself that serves as a $\beta$-function for the vacuum energy. We introduce therefore the RG-improved Coleman-Weinberg formula as follows:
\begin{equation}\label{RGCW}
{d \over d t} V_{CW}^F (H_c,t) = {N_c \over 16\pi^2} ( m_t^4(H_c, t) + \widetilde{m}_t^4 (H_c,t) ),
\end{equation}
where only the fermionic loops have been considered, the scalar loops giving contributions from the NNLL correction. In order to improve the potential up to the $t^2$ terms, we need to solve Eqs.~(\ref{BGBetaFunctions}) and find the renormalized top at twin top masses of Eq.~(\ref{MassesRen}) at the LL. The initial conditions for the Wilson coefficients are fixed at the scale $m_*$; in particular, the wave function $Z_{\widehat{\eta}}$ has the field-dependent starting value of Eq.~(\ref{ZEta}) and automatically re-sums the contribution to the Higgs mass induced by the higher-dimensional operators $\mathcal{O}_H$, $\mathcal{O}_H'$ and $\mathcal{O}_D$. After re-scaling the Higgs field to pass in the canonical basis,
\begin{equation}\label{Rescale}
H_c \rightarrow {H_c \over \sqrt{Z_H(t)}},
\end{equation}
we can derive $\lambda(t)$ and $c_6(t)$ at order $t^2$ from the RG-improved Coleman-Weinberg formula. We also need to compute the physical value of $c_H(t)$, which appears in the external correction of order $\xi$ to the Higgs mass:
\begin{equation}\label{cHRescaled}
c_H(t) \rightarrow {c_H(t)\over Z^2_H(t)} = c_H - {3 c_H y_t^2\over 8\pi^2}t.
\end{equation} 
Notice that $c_H'(t)$ does not receive contributions from the wave function of the Higgs field at order $t$ since it is only generated at one-loop. 

From Eq.~(\ref{HiggsMassFormula}), it is finally straightforward to find the IR correction to the Higgs boson mass at the NLL:
\begin{equation}\label{MHIRNLL}
\begin{array}{ll}
\displaystyle (M_H^2)_{IR}^{NLL}(t) =&\displaystyle {3 v^2\over 256\pi^4} \left[  \left( 16 g_S^2 y_t^4 + 16 \widetilde{g}_S^2 \widetilde{y}_t^4 -15 y_t^6 +3(c_H+1)\widetilde{y}_t^6-12y_t^2\widetilde{y}_t^4 \right) t^2\right. +\vspace{0.2cm}\\
& \displaystyle \left. \left( 36 c_H y_t^6 +\widetilde{y}_t^6 \left ( {9\over 8}  d_H - 12 c_H -12 c_H^2 - 6\right) -6 y_t^4 \widetilde{y}_t^2+24 c_H y_t^2\widetilde{y}_t^4 -  \right.\right. \vspace{0.2cm} \\
& \displaystyle  \left. \left. 16 c_H g_S^2   y_t^4  - 16 c_H \widetilde{g}_S^2\widetilde{y}_t^4 \right)~ \xi ~t^2\right].
\end{array}
\end{equation}
This is our final result for the model-independent RG evolution of the Higgs mass in a low-energy Twin Higgs theory. The renormalization scale $\mu$ encoded in the expansion parameter $t$ is taken to be a generic scale bigger than the physical twin top mass. When explicitly evaluating the Higgs mass, we will fix $\mu = m_t$ and match at the scale $\widetilde{m}_t$ where the twin tops need to be integrated out. Finally notice that the result in Eq.~(\ref{MHIRNLL}) agrees with the same solution derived with a more conventional diagrammatic approach in \cite{FuturePaper}. 

\section{The NNLL effective potential}
\label{sec:NNLL}

Since our Twin Higgs extension of the SM is a non-renormalizable theory, the RG-improvement of the Higgs effective potential is not completely exhausted by the $\beta$-functions we have just computed. These latter cannot capture all the physical effects coming into play at the next orders in $t$. Other higher-dimensional operators are in fact generated along the flow that contribute to the Higgs mass and that cannot be included in our previous background-dependent renormalization of the fermion masses. In order to fully capture the NNLL correction to the potential, we then need to classify a series of new quantum contributions to the twin top masses that are only present from the $t^3$ terms. Together with these effects, we have to take into account the RG-evolution of the strong couplings, whose running is negligible at the NLL order, and the scalar part of the Coleman-Weinberg potential. In this Section, we analyze the cubic correction to the low-energy action in the background field language studying in detail the contributions in each category. We will supplement the field-dependent $\beta$-functions with another set of RG-evolution equations for the twin top masses and solve them to systematically re-sum the leading logarithms. The expression of the Higgs mass at the NNLL order will be our final result.

\subsection{Running of the strong couplings and scalar contribution to the Coleman-Weinberg potential}

The first important correction to the NLL effective potential comes from the RG-evolution of the strong couplings, both in the SM and in the $\widetilde{\text{SM}}$. The Twin $\widetilde{SU}(3)$ strong interactions are an exact mirror copy of the $SU(3)$ gauge theory. They are both external to the whole mechanism that protects the Higgs mass from radiative corrections so that we can assume the $Z_2$ symmetry to be unbroken in this sector. The runnings of $g_S$ and $\widetilde{g}_S$ are therefore identical and both described by the standard QCD $\beta$-function with $n_f=6$ flavors. From our initial conditions at the scale $m_*$, we find:
\begin{equation}\label{gSRunning}
g_S(t) = g_S +{7 g_S^3 \over 32 \pi^2} t, \qquad \widetilde{g}_S(t) = \widetilde{g}_S + {7 \widetilde{g}_S ^3 \over 32 \pi^2} t,
\end{equation}
which give the strong couplings at the renormalization scale $\mu \ll m_*$.

The second non-trivial contribution comes from the scalar part of the Coleman-Weinberg potential, which re-sums all the vacuum energy loops involving the Higgs and the GB's. The generalization of Eq.~(\ref{RGCW}) is straightforward:
\begin{equation}\label{RGCWS}
{d\over dt } V_{CW}^S(H_c,t)= -{1\over 64 \pi^2}\left( \sum_{i=1}^3 (\widehat{m}^i_{GB})^4(H_c,t) + \widehat{m}_H^4(H_c,t) \right),
\end{equation}
where $\widehat{m}_{GB}^i$ and $\widehat{m}_H$ are respectively the masses of the quantum fluctuations for the three SM GB's and for the Higgs in the background field. They can be found by diagonalizing the mass term for the high-energy modes; from the general form of the potential in Eq.~(\ref{VPot}), in fact, after splitting as in Eq.~(\ref{Separ}), we find a non-diagonal mass matrix for $\widehat{\eta}$,
\begin{equation}\label{MassTerm}
\mathcal{L}_M(H_c) = - \widehat{M}^2_{ij}(H_c) \widehat{\eta}^{~i} ~\widehat{\eta}^{~j},
\end{equation}
where each of the $\widehat{\eta}^{~i}$ denotes a component of the full high-frequency doublet. The diagonalization of $\widehat{M}$ leads to the following expressions in the spectator background:
\begin{equation}\label{FlucMasses}
\begin{array}{ll}
\displaystyle (\widehat{m}_{GB}^1)^2= (\widehat{m}_{GB}^2)^2= (\widehat{m}_{GB}^3)^2 (H_c,t)= {1\over Z_{\widehat{\eta}}(H_c,t)}\left(\mu^2 (t)+ 8 \lambda(t) H_c^\dagger H_c + 24 c_6(t) {(H^\dagger_c H_c)^2\over f^2}\right),\vspace{0.2cm}\\
\displaystyle \widehat{m}_H^2(H_c,t) =  {1\over Z_{\widehat{\eta}}(H_c,t)}\left(\mu^2 (t)+24 \lambda(t) H^\dagger_c H_c +120  c_6(t) {(H^\dagger_c H_c)^2\over f^2}\right).
\end{array}
\end{equation}
The presence of the wave function for $\widehat{\eta}$ is again a feature of our non-canonical basis. When finding the masses for the physical fields, we need to redefine the fluctuation thus getting an explicit dependence from $Z_{\widehat{\eta}}$ in the scalar masses.

The correction to the low-energy action from the scalar Coleman-Weinberg potential can only arise at cubic order in the logarithmic expansion. This is because $\lambda$ in our theory is first generated at one-loop, so that when integrating Eq.~(\ref{RGCWS}) we cannot find a lower contribution. For the NNLL result, we do not need to compute $c_6$, since it gives an effect suppressed by $\xi$. We reported, however, the full expression of the scalar masses for completeness. Finally, also in the scalar sector, the computation of the running of the Higgs quartic coupling through the background field method is perfectly equivalent to the diagrammatic approach. At the NNLL, it is in one-to-one correspondence only with the one-loop diagram generated by the Higgs self-interaction. With the background technique, however, one has the advantage to avoid deriving any symmetry factor, that can be cumbersome in the standard procedure. 

\subsection{Renormalization of the twin top mass in the Higgs background}

The second class of effects that contribute to the Higgs mass at the NNLL order is related to the renormalization of the twin top mass induced by the non-linear interactions between the quarks and the scalar fluctuation and by new higher-dimensional operators. Let us start considering how the twin propagator is affected by the non-linear coupling with $\widehat{\eta}$. After splitting the high-energy modes from the long-distance degrees of freedom, the Lagrangian in Eq.~(\ref{LagF}) develops an additional background-dependent quadratic interaction as follows:
\begin{equation}\label{LagFQuad}
\mathcal{L}^F(m_*) \supset {\widehat{\widetilde{y}}_2^{~GB}(H_c) \over 2 \sqrt{2} f} \sum_{i=1}^3 ~\bar{\widetilde{t}}~\widetilde{t} ~\widehat{\eta}_i^{~2} +  {\widehat{\widetilde{y}}_2^{~H}(H_c) \over 2 \sqrt{2} f}  ~\bar{\widetilde{t}}~\widetilde{t} ~\widehat{\eta}_4^{~2}.
\end{equation} 
In the previous equation, we have explicitly written the quantum fluctuation in components,
\begin{equation}\label{EtaComp}
\widehat{\eta} ={1\over \sqrt{2}} \left( 
\begin{array}{ll}
\widehat{\eta}_1 + i \widehat{\eta}_2 \\
\widehat{\eta}_4 + i \widehat{\eta}_3
\end{array}\right),
\end{equation}
indicating with $\eta_4$ the high-energy modes of the physical Higgs and with the remaining $\eta_i$ those of the three GB's. The twin tops interact  differently with the various types of scalar fluctuations and we have introduced two field-dependent couplings:
\begin{equation}\label{QuadCoup}
\widehat{\widetilde{y}}_2^{~GB} (H_c) = {\widetilde{y}_t \over \sqrt{1-{2H^\dagger_c H_c\over f^2}}}, \qquad  \widehat{\widetilde{y}}_2^{~H} (H_c) = {\widetilde{y}_t \over \left(1-{2H^\dagger_c H_c\over f^2}\right)^{3/2}}.
\end{equation}
The first one denotes the interaction with the three GB's, which are all coupled identically with fermions. The physical Higgs, instead, picks up an additional term after expanding the doublet and it is coupled differently with respect to the other scalars.  

\begin{figure}[t!]
\begin{center}
\includegraphics[width=0.9\textwidth] {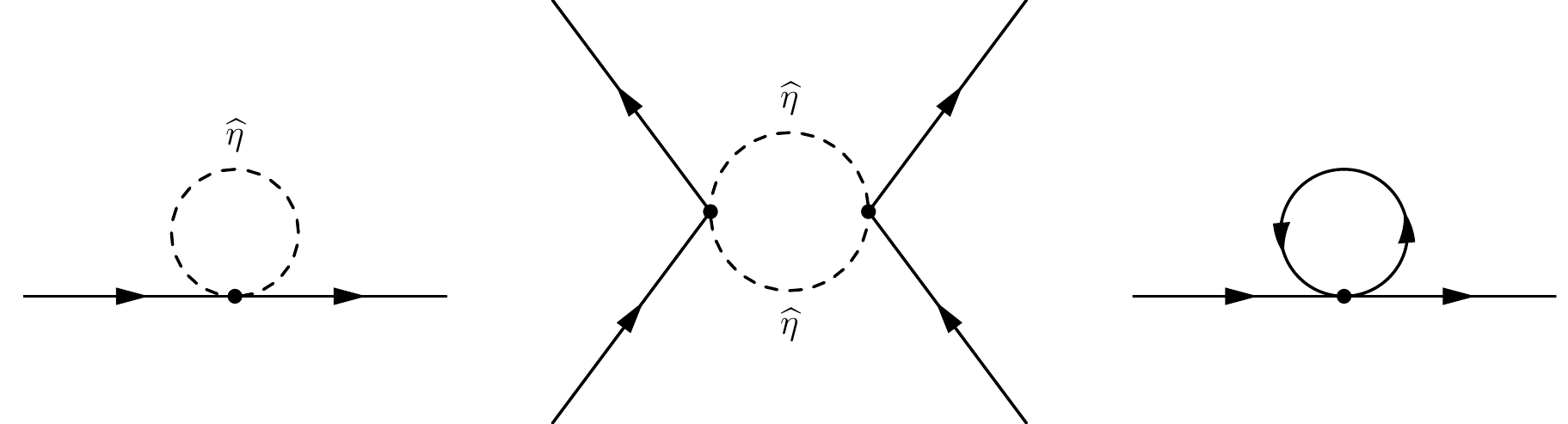}
\end{center}
\caption{One-loop diagrams displaying the renormalization of the twin top quark mass. On the left, the diagram correcting the twin top propagator with loops of scalars; in the middle the one generating the four-fermion operator of Eq.~(\ref{FourFermion}); on the right, the renormalization of the twin top propagator due to the four-fermion interaction. Solid lines indicate the twin quarks, dashed lines the scalar fluctuation.}
\label{fig:FourFermion}
\end{figure}

The existence of these quadratic interactions induces a renormalization of the twin top propagator due to scalar tadpoles, as in Fig.~(\ref{fig:FourFermion}). In particular, no correction to the fermion wave function can be generated and we find only a quantum contribution to the twin mass:
\begin{equation}\label{RinMass1}
\delta \widetilde{m}_t^S(H_c,t) = 3 {\widehat{\widetilde{y}}_2^{~GB}(H_c) \over 32\sqrt{2}\pi^2} {(\widehat{m}^1_{GB})^2(H_c,t) \over f} t +{\widehat{\widetilde{y}}_2^{~H}(H_c) \over 32 \sqrt{2}\pi^2} {\widehat{m}_{H}^2(H_c,t) \over f }t.
\end{equation}
The renormalization of $\widetilde{m}_t$ is proportional to the field-dependent scalar masses, which originate first at the LL. The correction to the Higgs effective potential must then arise at cubic order, as expected.

We consider now the class of physical effects due to the generation of new higher-dimensional operators that are not captured by the field-dependent $\beta$-functions of the top Yukawas. The first of these operators is the six-dimensional four-fermions interaction obtained by integrating out the high-frequency scalar modes, as shown in the median diagram of Fig.~(\ref{fig:FourFermion}). At one-loop, the Lagrangian in the fermionic sector receives the following additional contribution:
\begin{equation}\label{FourFermion}
\mathcal{L}^F(t) \supset {c_{4t}(H_c, t) \over 4 f^2} \left(\bar{\widetilde{t}}~\widetilde{t}\right)^2,
\end{equation}
with
\begin{equation}\label{C4t}
c_{4t}(H_c,t) = 3 {\widehat{\widetilde{y}}_2^{~GB} (H_c)^2\over 16\pi^2 Z_{\widehat{\eta}}^2(H_c)} t +  {\widehat{\widetilde{y}}_2^{~H} (H_c)^2\over 16\pi^2 Z_{\widehat{\eta}}^2(H_c)} t.
\end{equation}
In the background field language, this operator affects the Higgs potential by renormalizing the twin top propagator, as it can be seen again in the last diagram of Fig.~(\ref{fig:FourFermion}). It is straightforward to derive a second correction to the fermion mass which reads:
\begin{equation}\label{RinMass2}
\delta\widetilde{m}_t^F (H_c,t) = - {N_c \over 4 \pi^2} c_{4t}(H_c,t) {\widetilde{m}_t^3 (H_c)\over f^2} t.
\end{equation}
The joint quantum correction to the four fermion interaction and to the twin mass implies a contribution to the low-energy action only at NNLL.

There is a second kind of higher-dimensional operators renormalizing the twin top mass which are seeded along the flow by $\mathcal{O}_H$, $\mathcal{O}_H'$ and $\mathcal{O}_D$ and which are distinct from the ones captured by the wave function renormalization of $\widehat{\eta}$. After splitting the high-energy modes from the low-energy degrees of freedom, in fact, not only do those operators induce a non-canonical kinetic term for $\widehat{\eta}$, but they also generate other interactions involving derivatives of the external background field. These latter were previously neglected since their contribution to the Higgs mass is first encountered at the NNLL. For instance, according to the notation of \cite{SILH2}, in the SM sector one would get at one-loop the current-current operators $\mathcal{O}_L^t = i (H_c^\dagger \overleftrightarrow{D}_\mu H_c) \bar{Q}_L \gamma^\mu Q_L$, $\mathcal{O}_L^{(3)t} = i (H_c^\dagger \sigma^a \overleftrightarrow{D}_\mu H_c) \bar{Q}_L \gamma^\mu \sigma^a Q_L$ and $\mathcal{O}_R^t = i (H_c^\dagger \overleftrightarrow{D}_\mu H_c) \bar{t}_R \gamma^\mu t_R$. These latter can only renormalize the effective potential at order $\xi $, since they contribute to the running of $c_6$. They therefore do not belong to the NNLL order and we neglect them. Analogous current-current operators in the $\widetilde{\text{SM}}$ sector cannot be generated. The Higgs currents $H_c^\dagger \overleftrightarrow{D}_\mu H_c$ and $H_c^\dagger \sigma^a \overleftrightarrow{D}_\mu H_c$ transform in fact as a $({\bf{3}},{\bf{1}})$ and a $({\bf{1}},{\bf{3}})$, respectively, under the custodial group $SO(4) \sim SU(2)_L \times SU(2)_R$, whereas the twin tops are global singlets under this symmetry. In the SM, the Yukawa coupling transforms as a $({\bf{2}},{\bf{1}})$; it is then possible to form an $SU(2)_L$ total singlet proportional to $y_t^2$ and the current-current operators are allowed by selection rules. In the twin sector, these latter are instead forbidden by the quantum numbers, since the twin top Yukawa transforms as $({\bf{1}},{\bf{1}})$; an operator of the type $i(H_c^\dagger \overleftrightarrow{D}_\mu H_c)\bar{\widetilde{t}}\gamma^\mu \widetilde{t}$ is therefore absent because of selection rules.

\begin{figure}[t!]
\begin{center}
\includegraphics[width=0.9\textwidth] {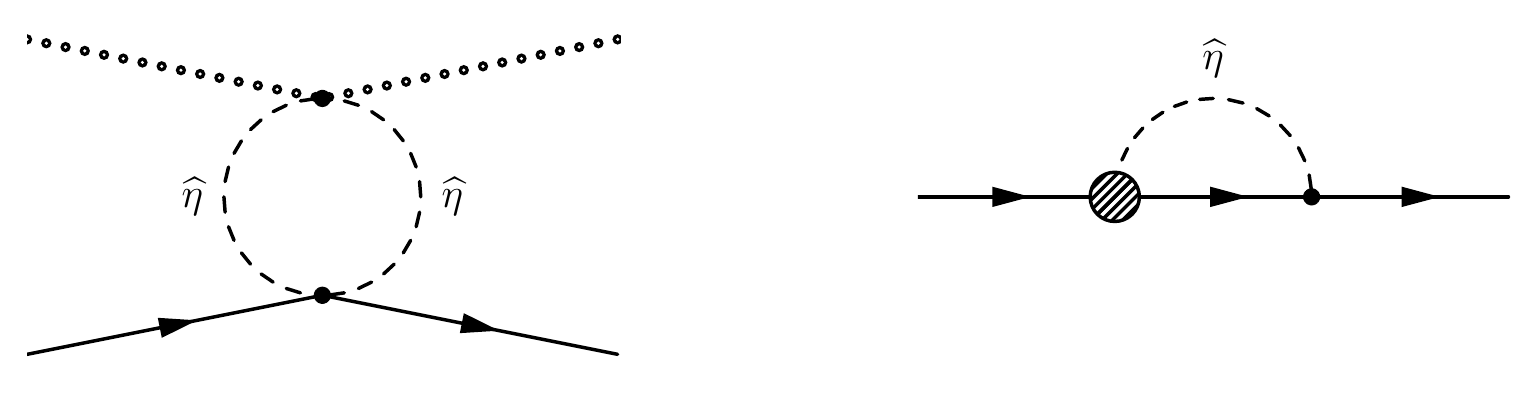}
\end{center}
\caption{The one loop-diagrams displaying the generation of the operator $\mathcal{O}_{\Box}$ (on the left) and the renormalization of the twin top mass (one the right). The blob in the last diagram denotes insertions of $\mathcal{O}_{\Box}$. The external dotted lines indicate the background field, the internal dashed ones the dynamical fluctuation; the solid lines indicate again the twin tops.}
\label{fig:CurrentOperator}
\end{figure}

The only type of higher-dimensional operator involving derivatives of the external field that is generated in the twin top sector has dimension seven and is of the form:
\begin{equation}\label{BoxOperator}
\mathcal{O}_\Box = - H_c^\dagger \Box H_c ~ \bar{\widetilde{t}} ~ \widetilde{t} + \text{h.c.}.
\end{equation}
It is made up of total singlets and is allowed by the symmetries of our theory. From the original Lagrangian (\ref{LagInitial}), after the redefinition in Eq.~(\ref{Separ}), one finds the following interaction between the scalar fluctuations and the background field that seeds exactly this operator:
\begin{equation}\label{LEtaBox}
\mathcal{L}(m_*) \supset -{1\over 2 f^2} \left( 2c_H + c_H' + d_H{ H_c^\dagger H_c\over 2 f^2} \right) \left( \eta_4 ^2 + \sum_{i=1}^3\eta_i^2\right) H_c^\dagger \Box H_c + \text{h.c.}.
\end{equation}
After integrating out the short-distance degrees of freedom, we can generate at order $t$ the following contribution to the Lagrangian,
\begin{equation}\label{LBox}
\mathcal{L}(t) \supset {c_\Box (H_c, t) \over f^3}\mathcal{O}_\Box;
\end{equation}
the background-dependent Wilson coefficient is obtained by computing the diagram on the left in Fig.~(\ref{fig:CurrentOperator}). We find:
\begin{equation}\label{cBox}
c_\Box(H_c, t ) = \left( 2 c_H + c_H'+d_H {H_c^\dagger H_c\over 2 f^2}\right) {\left ( 3\widehat{\widetilde{y}}^{~GB}_2(H_c)+ \widehat{\widetilde{y}}^{~H}_2(H_c)\right) \over 16 \sqrt{2}\pi^2 Z_{\widehat{\eta}}^2(H_c) }  t.
\end{equation}
Notice that $c_H'$ is zero at the scale $m_*$ and it is first generated at one-loop, so that it will not give a contribution to the NNLL effective potential through the operator $\mathcal{O}_\Box$. We reported its correction to $c_\Box$ for completeness.

The operator $\mathcal{O}_\Box$ contributes to the Higgs potential by renormalizing the twin top mass, as depicted in the last diagram of Fig.~(\ref{fig:CurrentOperator}). We formally need to split the high-energy modes a second time and keep only the interactions with the box operator acting on the fluctuating field. We have:
\begin{equation}\label{BoxSplit}
\mathcal{L}(t) \supset -\left(c_\Box(H_c,t) {H^\dagger_c \over f}\right) {\Box\widehat{\eta}~\bar{\widetilde{t}} ~ \widetilde{t}\over f^2} + \text{h.c.}.
\end{equation}
The field-dependent correction to the twin masses is then found to be:
\begin{equation}\label{RinMass3}
\delta \widetilde{m}_t^\Box (H_c,t) = - c_\Box(H_c,t) {2H^\dagger_c\over f} {\widehat{\widetilde{y}}_t(H_c) \over 8\sqrt{2}\pi^2 Z_{\widehat{\eta}}(H_c)} {\widetilde{m}_t^3(H_c)\over f^2} t ,
\end{equation}
where we used the notation of Eq.~(\ref{Yukawas}) for the coupling $\widehat{\widetilde{y}}_t(H_c)$. Together with the previous two quantum contributions, this formula gives the last renormalization of the fermion masses entering the effective action at the NNLL order.

We finally summarize the results obtained in this Section with a set of $\beta$-functions for the higher-dimensional operators and the twin top masses. They will supplement the evolution equations we already have for the Yukawa couplings and re-sum all the leading logarithms appearing in the Higgs mass. From our previous expressions, we immediately find:
\begin{equation}\label{RGBetaFunctionDue}
\begin{array}{ll}
\displaystyle \beta_{c_{4t}} = {\left(  3 \widehat{\widetilde{y}}_2^{~GB}(H_c,t)^2+  \widehat{\widetilde{y}}_2^{~H}(H_c,t)^2\right)\over 16 \pi ^2 Z_{\widehat{\eta}}^2(H_c,t)},\vspace{0.2cm}\\
\displaystyle  \beta_{c_\Box} = \left( 2 c_H(t) + c_H'(t) +d_H {H_c^\dagger H_c\over 2 f^2}\right) {\left ( 3\widehat{\widetilde{y}}^{~GB}_2(H_c,t)+ \widehat{\widetilde{y}}^{~H}_2(H_c,t)\right) \over 16 \sqrt{2}\pi^2 Z_{\widehat{\eta}}^2(H_c,t) },\vspace{0.2cm}\\
\displaystyle \beta_{\widetilde{m}_t^S}= {3\widehat{\widetilde{y}}_2^{~GB}(H_c,t) (\widehat{m}_{GB}^1)^2(H_c,t)+ \widehat{\widetilde{y}}_2^{~H}(H_c,t) \widehat{m}_{H}^2(H_c,t)\over 32\sqrt{2}\pi^2 f} ,\quad \beta_{\widetilde{m}_t^F} = -{3\over 4\pi^2} c_{4t}(H_c,t){\widetilde{m}_t^3(H_c,t) \over f^2},\vspace{0.2cm}\\
\displaystyle \beta_{\widetilde{m}^\Box_t} = - c_\Box(H_c,t) {2H^\dagger_c H_c\over f^2}{\widetilde{y}_t(H_c,t)\over \sqrt{1-{2H_c^\dagger H_c\over f^2}}} {1\over 8 \sqrt{2}\pi^2 Z_{\widehat{\eta}}(H_c,t)} {\widetilde{m}_t^3(H_c,t)\over f^2}.
\end{array}
\end{equation}
The quadratic couplings $\widehat{\widetilde{y}}_2^{~GB}$ and $\widehat{\widetilde{y}}_2^{~H}$ acquire in general a dependence on the expansion parameter through the evolution of the twin Yukawa. The background-dependent twin top mass at a generic order in $t$ is now defined as:
\begin{equation}\label{MassesRen2}
\widetilde{m}_t (H_c,t) = {\widetilde{y}_t(H_c, t) f\over \sqrt{2}}\sqrt{1-{2 H_c^\dagger H_c\over f}}+ \widetilde{m}_t^S(H_c,t) + \widetilde{m}_t^F(H_c,t)+ \widetilde{m}_t^\Box(H_c,t),
\end{equation}
where the last three additional terms correspond to the solution of the previous $\beta$-functions in the Higgs spectator field. This formula together with the RG equations are the basic elements to compute the Higgs potential at NNLL.

\subsection{Higgs mass at the NNLL}

In order to find the Higgs effective potential at the NNLL, we solve the $\beta$-functions in Eqs.~(\ref{BGBetaFunctions}) and (\ref{RGBetaFunctionDue}) up to order $t^2$ and use Eqs.~(\ref{MassesRen}) and (\ref{MassesRen2}) to derive the renormalized background-dependent fermionic masses. Adding the running of the strong couplings and the scalar contribution in Eq.~(\ref{RGCWS}), we have in the canonical basis:
\begin{equation}\label{MHIRNNLL}
\begin{array}{ll}
\displaystyle (M_H^2)_{IR}^{NNLL}(t) =&\displaystyle {v^2\over 8192\pi^6} \left[  736 g_S^4 y_t^4 - 1104 g_S^2 y_t^6+387 y_t^8 + \widetilde{y}_t^4\left( 736 \widetilde{g}_S^4 -288 g_S^2 y_t^2- 576 \widetilde{g}_S^2 y_t^2+\right.  \right. \vspace{0.2cm}\\
& \displaystyle \left. \left. 18 (3-4 c_H)y_t^4\right) + \widetilde{y}_t^6 \left(240(1+c_H)\widetilde{g}_S^2-18(7+8 c_H) y_t^2  \right)  - \right. \vspace{0.2cm}\\
& \displaystyle \left. 16 \widetilde{y}_t^8 \left(18 +37 c_H + 30 c_H^2-{11\over 4} d_H  \right) \right] ~t^3.
\end{array}.
\end{equation}
All the parameters are again evaluated at $m_*$, which sets the scale where the RG-evolution of the Wilson coefficients starts. 

\section{Results}
\label{sec:Results}

The background field method proved to be a useful technique to automatically re-sum a whole series of diagrams, compute the renormalized effective potential and derive an expression for the Higgs mass valid up to the NNLL. Our final prediction for the IR RG-evolution of this observable is the sum of three different contributions:
\begin{equation}\label{MHIRTot}
(M_H^2)_{IR}(\mu) = (M_H^2)^{LL}_{IR}(\mu)+ (M_H^2)^{NLL}_{IR}(\mu)+ (M_H^2)^{NNLL}_{IR}(\mu),
\end{equation}
which are given respectively in Eqs.~(\ref{MHIRLL}), (\ref{MHIRNLL}) and (\ref{MHIRNNLL}). The renormalization scale $\mu$ is encoded in the expansion parameter $t=\log(m_*^2/ \mu^2)$ and is chosen to be the energy scale where the Higgs mass is measured, for instance the top mass. From our analytic result, we can now obtain a numerical estimate of $(M_H^2)_{IR}(m_t)$ and compare it with the experimental observations. This in turn will give us an idea of the capability of the low-energy Twin Higgs construction to predict the Higgs mass in the correct range only through the IR physics. We will also try to estimate the UV correction that would be needed in order to match with experiments. The prediction of the Higgs mass at cubic precision is therefore an important test of the Twin Higgs scenario as a new paradigm for understanding physics at the EW scale.

In order to derive a numerical estimate of the Higgs mass, we have first to assign a value to all the Wilson coefficients appearing in the final formula. The initial conditions for their RG-evolution are fixed at the scale $m_*$; we know already that $c_H=1$ and $d_H=8$ due to the pNGB nature of the Higgs field. Because of Twin parity, which is still approximately a good symmetry at $m_*$, we can set $\widetilde{g}_S=g_S$ and $\widetilde{y}_t = y_t$; the strong and the Yukawa couplings, however, are measured at the IR scale $m_t$ and we must solve their RG evolution equation to run their value up to the UV. We need to derive $g_S$ at first order in the logarithmic expansion, whereas $y_t$ must be known up to the quadratic contributions. We have:
\begin{equation}\label{PhysicalCouplings}
\begin{array}{ll}
\displaystyle g_S = \widetilde{g}_S = g_E- {7 g_E^3\over 32\pi^2}\log\left({m_*^2\over m_t^2} \right),\vspace{0.2cm}\\
\displaystyle y_t = \widetilde{y}_t = y_E+{y_E (9 y_E^2-16 g_E^2)\over 64 \pi^2}\log\left({m_*^2\over m_t^2} \right)+{y_E(704 g_E^4 - 576 g_E^2 y_E^2 + 243 y_E^4)\over 8192 \pi^4} \log^2\left({m_*^2\over m_t^2} \right),
\end{array}
\end{equation}
where $y_E$ and $g_E$ indicate the experimental value of these couplings at the scale $m_t$. For the Yukawa, we use the $\overline{\text{MS}}$ value of the top quark mass, $m_t^{\overline{\text{MS}}} = 160~ \text{GeV}$, from which we derive $y_E \sim 0.92$. For the strong interaction, we run the parameter measured at the scale of the $Z$ boson mass, $g_S(M_Z) \sim 1.22$, to the top mass scale, so we have $g_E \sim 1.17$. Notice that the RG evolution of the top Yukawa in Eq.~(\ref{PhysicalCouplings}) coincides with the solution of the $\beta$-functions in Eqs.~(\ref{BGBetaFunctions}) for vanishing external field after re-scaling the Higgs spectator as in Eq.~(\ref{Rescale}).

The last aspect we must take care of when estimating the Higgs mass is the existence of the twin top mass threshold. We have previously derived all our results at a generic scale $\mu \gg \widetilde{m}_t$; if we want to fix $\mu = m_t$, we need to integrate out the Twin partners at the scale $\widetilde{m}_t$ and resume the purely SM running from this scale down to the top quark mass. Our Higgs mass is then the sum of two pieces: a first evolution from $m_*$ to $\widetilde{m}_t$ which serves as the initial condition for a second contribution from $\widetilde{m}_t$ to $m_t$. This latter is obtained by switching off the twin parameters and keeping only the SM supplemented by dimension-six operators. The twin mass is evaluated at the scale $\widetilde{m}_t$ using Eq.~(\ref{MassesRen2}), setting the external background to its physical vacuum expectation value and expanding at first order in $\xi$. 

\begin{figure}[t!]
\begin{center}
\includegraphics[width=0.9\textwidth] {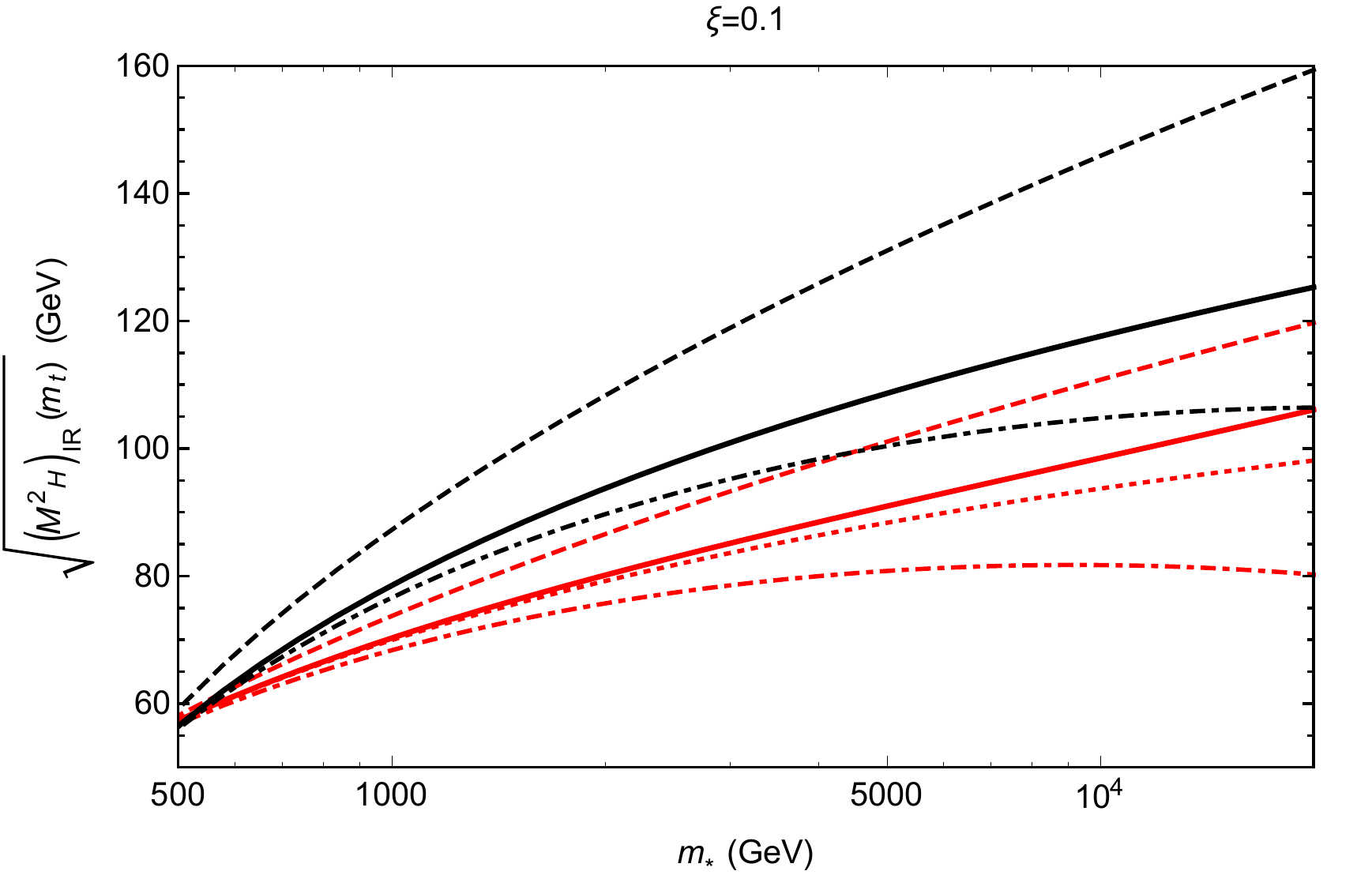}
\end{center}
\caption{\small IR contributions to the Higgs mass in logarithmic scale, both in the full Twin Higgs theory and in the pure SM: LL contribution (dashed black curve),  NLL contribution (dashed dotted black curve), NNLL contribution (thick black curve), LL SM contribution (dashed red curve), NLL SM contribution (dashed dotted red curve), NNLL SM contribution (thick red curve), re-summed total SM contribution (dotted red curve).}
\label{fig:HiggsMass}
\end{figure}

\begin{figure}[t!]
\begin{center}
\includegraphics[width=0.5\textwidth] {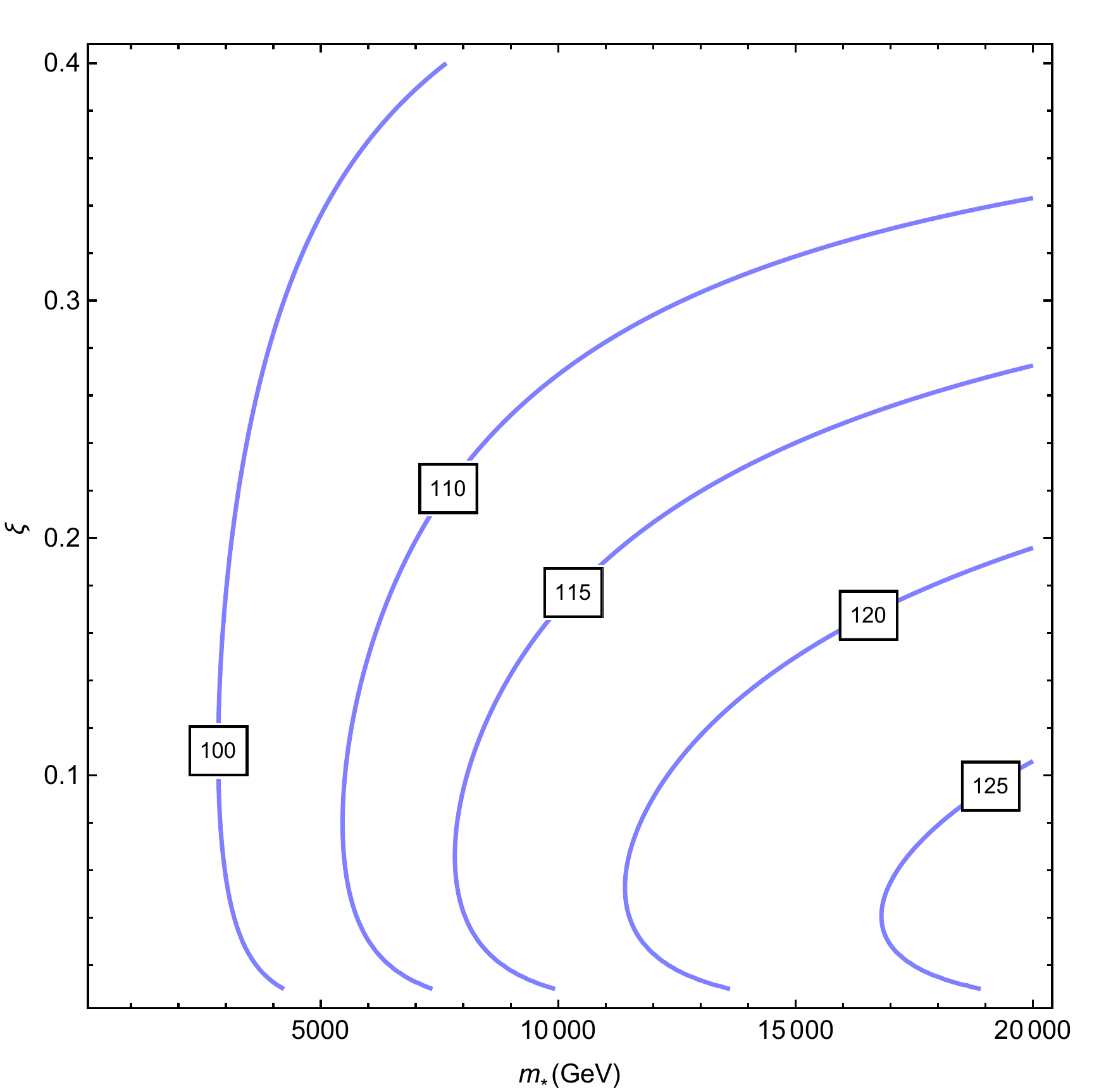}
\end{center}
\caption{\small Contour plots of the renormalized Higgs mass (in GeV) at NNLL in the plane $(m_*, \xi)$.}
\label{fig:HiggsMass2}
\end{figure}

Our final results are shown in Fig.~(\ref{fig:HiggsMass}), where we plot the value of the Higgs mass at the scale $m_t$ as function of the cut-off $m_*$ for the fixed value of $\xi=0.1$. We choose this latter in agreement with the general constraint due to EWPT. Fig.~(\ref{fig:HiggsMass}) shows two different sets of curves, a first one in black for the full prediction in the Twin Higgs low-energy model and a second one in red for the pure SM quartic coupling evolution. In each of the two cases, we reported the Higgs mass at the LL, the NLL and the NNLL. For both results, the LL solution appears to be quite an overestimation of the logarithmic series, indicating the importance of extending the computation to the higher orders including the effects of the top Yukawa running. At the NLL, the Higgs mass reduces drastically because $y_t$ and $\widetilde{y}_t$ become considerably smaller along the flow from $m_t$ to $m_*$ due to QCD effects. For the Twin Higgs model we get $(M_H^2)_{IR}^{NLL} (m_t)\sim (105~ \text{GeV})^2$ with a cut-off at $10-20 ~\text{TeV}$, which is considerably bigger than the SM value of $(80 ~ \text{GeV})^2$ due to the presence of the extra light degrees of freedom. The truncation of the logarithmic series to quadratic order, however, is still a rude approximation of the re-summed solution; we see in fact that the NNLL introduces non-negligible effects already for $m_* \sim 2 - 3~ \text{TeV}$ and for bigger values of the cut-off the NLL solution becomes less reliable. At cubic order, the prediction for the Higgs mass increases in both cases, mostly due to QCD effects that tend to rise the value of $y_t$, as in Eq.~(\ref{PhysicalCouplings}), and of its corresponding twin. The growth of $M_H^2$ in the Twin Higgs model is however less sharp than in the SM, because of non-renormalizable effects. In particular, the contributions to the effective potential from four fermions interactions and from the operator $\mathcal{O}_\Box$ are both negative and tend to reduce the Higgs mass with respect to QCD. We may wonder if the NNLL solution is a reliable approximation for values of the cut-off scale of $10-20~\text{TeV}$ or if quartic effects will still give non-negligible corrections. We do not have a result at this order in the logarithmic series for the Twin Higgs model, but we can estimate its behavior studying the SM. We reported in Fig.~(\ref{fig:HiggsMass}) also the re-summed SM solution for the Higgs mass obtained after solving numerically the $\beta$-function for the quartic coupling. The comparison of this latter with the NNLL prediction shows that the cubic approximation in the SM can be considered reliable up to $m_* \sim 20~ \text{TeV}$, for which value the difference between the two solutions is indicatively 5\%. We can expect that something similar will happen also in the Twin Higgs case. Despite the presence of non-renormalizable corrections, in fact, the QCD effects are still the dominant ones and they must behave exactly as in the SM. The full solution must then decrease with respect to the NNLL correction and we expect our NNLL solution to be a reliable approximation for $m_* \sim 20 ~\text{TeV}$. Beyond this value for the cut-off, the quartic contributions must necessarily be taken into account and our computation cannot be trusted any longer. 

After discussing the validity of our approximation, we can now specifically consider the prediction of the Higgs mass that we get in the Twin Higgs model up to the NNLL order. From Fig.~(\ref{fig:HiggsMass}), we see that $(M_H^2)_{IR}^{NNLL} \sim 120 ~\text{GeV}$ for $m_* \sim 10 - 20 ~\text{TeV}$, a value which is in the perfect range to match with the experimental observations, $(M_H^2)_{Exp} = 125 ~(\text{GeV})^2$. We also show in Fig.~(\ref{fig:HiggsMass2}) the contour plots for the renormalized Higgs mass at NNLL in the plane $(m_*, \xi)$, so as to visualize the effects of the fine-tuning parameter as the cut-off scale changes. We find again that with a moderate tuning, $\xi \sim 0.1-0.2$, and a value of $m_*$ around 20 TeV it is possible to reproduce the experimental results. The IR physics alone can therefore generate an acceptable value for the Higgs mass through the RG-evolution. The remaining part that is missing to agree with observations could be supplemented by a small UV contribution. For example, with $\xi =0.1$ and $m_* \sim  10 ~ \text{TeV}$, a value of the cut-off for which our computation is more reliable, a modest $(M_H^2)_{UV} \sim (5 ~\text{GeV})^2$ is enough for the Twin Higgs paradigm to be matched perfectly with experiments. The smallness of the UV effect together with the possibility of pushing $m_*$ up to $\sim 20~ \text{TeV}$ are also necessary for the whole mechanism to make sense. On one side, the fact the $(M_H^2)_{UV}$ can be small confirms that the Higgs boson is not sensitive to the UV physics. On the other side, if $m_*$ can be very large, in the multi-TeV range, it is reasonable to neglect all the tree-level initial conditions for the higher-dimensional operators generated after integrating out the new physics. Their Wilson coefficients at the scale $m_*$ are model-dependent and suppressed by inverse powers of the cut-off; we expect them to give only a very small contribution to the Higgs mass. It is therefore approximately correct to set them to zero at $m_*$ and consider only their one-loop value seeded by the six-dimensional operator present at the tree-level, $\mathcal{O}_H$, and automatically captured by the background field method. Our prediction is then consistently model-independent and results only from the IR physics.

The lesson we can learn from the Twin Higgs mechanism is that it is possible to construct models with a natural light Higgs in the spectrum without necessarily requiring the existence of new light colored top partners. The Higgs can be insensitive to the UV scale of the heavy resonances charged under the SM, which can pushed up to $\sim 20~\text{TeV}$ for the experimental value of $M_H$ to be almost exactly reproduced by the IR physics through RG effects. The UV contribution must be small and any UV completion that can be imagined must be able to generate a modest value of the quartic coupling at the cut-off scale. Composite UV completions, for instance, can be easily realized that fulfill this requirement, \cite{CompositeTwinHiggs}.

\section{Conclusions}
\label{sec:Conclusions}

In this paper, we have computed the RG-improved Higgs effective potential and mass in the Twin Higgs model up to third order in logarithmic accuracy. We have carried out the calculation in the most general setting, writing an effective Lagrangian comprising only the IR degrees of freedom, namely the Higgs doublet, the SM quarks and their twins. In this way, our prediction for the Higgs mass is completely model-independent and proper to any possible UV completion, supersymmetric or composite, of the Twin Higgs paradigm. We have discussed the validity of our approximation. First of all, the Higgs potential is insensitive to the UV physics and we expect that the most important contributions to the mass come from the RG evolution due to loops of the IR degrees of freedom. Secondly, we have neglected the initial conditions for the higher-dimensional operators generated at the tree level after integrating out the UV physics. Their Wilson coefficients at $m_*$ are in fact suppressed by the weak coupling between the elementary and the UV sectors as well as by inverse powers of the cut-off $m_*$, which is reasonably of the order of $10-20 ~\text{TeV}$. Their contribution to the running of the potential, which is model-dependent, is therefore safely negligible. 

We showed how to carry out the renormalization of the potential in the most efficient way using the background field method. This technique proved to be extremely useful in order to re-sum the one-loop diagrams contributing to the running without necessarily classifying all the non-renormalizable operators in the Twin sector. We applied this method to our low-energy Lagrangian and we systematically included all the physical effects that are relevant up to the NNLL order. The final result can be obtained by solving a simple set of background-dependent $\beta$-functions from which we find the top and twin top masses in the spectator field. The Coleman-Weinberg formula for the effective action can then be easily applied to derive the Higgs mass at cubic order in the logarithmic expansion.

Our final prediction for the Higgs mass is summarized in Figs.~(\ref{fig:HiggsMass}) and (\ref{fig:HiggsMass2}) where we plot this observable as a function of the cut-off of the theory. At the NNLL, we get a value of the order of $M_H \sim 120 ~\text{GeV}$ for $m_* \sim 10-20 ~\text{TeV}$, which is in the perfect range to match with the experimental observations without requiring a big UV contribution. The IR degrees of freedom are then enough to account for the measured value of the Higgs mass through the RG-evolution of the effective potential. 

The background field computation developed in this paper can be improved in order to re-sum the whole logarithmic series and possibly get a numerical solution valid at all orders in the expansion parameter. For this purpose, a classification of the operators in the Twin sector seems unavoidable. Writing in full generality the Lagrangian at the scale $m_*$ including all the possible gauge invariant operators, we can again split the low-energy degrees of freedom from the short distance modes and integrate out these latter. After computing the most general $\beta$-functions for the running of the top mass and its twin, one could easily find the potential using the RG-improved Coleman-Weinberg formula without needing to specify which operators are generated at each order in the expansion in logarithms. From this point of view, our application of the background field method is not the most efficient one, since we had to understand for the NNLL solution which operators we expected to produce at one-loop that were not already captured in the background-dependent $\beta$-functions for the NLL result. Such a procedure would make it even more cumbersome to compute the quartic correction to the effective potential, because one would need to separately derive the evolution of $\mathcal{O}_D$, for instance, and again individuate the operators that were not previously included and that could give new contributions to the Higgs mass. A complete classification of the gauge invariant operators in the Twin sector together with the six-dimensional ones already listed for the SM would then provide the best way to systematically apply the background field method to the Twin Higgs model. This could be a very interesting extension of our results and could give more information on the stability of the effective potential with respect to the UV physics. 

\section*{Acknowledgements}
We thank Riccardo Rattazzi, Roberto Contino, Riccardo Torre and Rakhi Mahbubani for useful discussions. D.G. is supported by the Swiss National Science Foundation under contract 200020-150060. 

\providecommand{\href}[2]{#2}

\end{document}